\documentclass{entcs} 

\usepackage{arxivmacro}
\usepackage{times}
\usepackage{url}
\usepackage{lhs2TeXPreamble}
\usepackage{graphics}

\def\lastname{Kort \& L{\"a}mmel}

\newcommand{\mysubsection}[1]{\subsection{\textbf{#1}}}
\renewcommand{\qquad}{\hspace{50\in}}

\begin{document}

\begin{frontmatter}

\title{A Framework for Datatype Transformation\vspace{-77\in}}

\author{Jan Kort\,${}^1$ and Ralf L{\"a}mmel\,${}^{2,3}$}
\address{$^1$\,Universiteit van Amsterdam\\
$^2$\,Centrum voor Wiskunde en Informatica\\
$^3$\,Vrije Universiteit van Amsterdam\\\vspace{-100\in}
}

\begin{abstract}
We study one dimension in program evolution, namely the evolution of
the datatype declarations in a program. To this end, a suite of basic
transformation operators is designed. We cover structure-preserving
refactorings, but also structure-extending and -reducing
adaptations. Both the object programs that are subject to datatype
transformations, and the meta programs that encode datatype
transformations are functional programs.
\end{abstract}

\end{frontmatter}


\section{Introduction}

\vspace{-22\in}

\noindent
We study operators for the transformation of the datatype declarations
in a program. The presentation will be biased towards the algebraic
datatypes in Haskell, but the concepts are of relevance for many typed
declarative languages, e.g., Mercury and SML, as well as frameworks
for algebraic specification or rewriting like ASF+SDF, CASL, Elan, and
Maude.  Our transformations are rather syntactical in nature as
opposed to more semantical concepts such as data refinement. Our
transformations contribute to the more general notion of
\emph{functional program refactoring}~\cite{TR01}.

\smallskip

\noindent
The following introductory example is about extracting a new datatype
from constructor components of an existing datatype. This is
illustrated with datatypes that represent the syntax of an imperative
language. The following extraction identifies a piece of syntax to 
enable its reuse in later syntax extensions:

\smallskip

{\footnotesize
\input{snip/little-fold.math}
}

\noindent
In the present paper, we describe the design of a framework for
datatype transformations including the operators for the above
extraction. In Sec.~\ref{S:concerns}, we identify all the concerns
addressed by the framework. In Sec.~\ref{S:suite}, we describe all the
basic operators for datatype transformations. In Sec.~\ref{S:meeting},
these operators are lifted from datatypes to complete
programs. Related work is discussed in Sec.~\ref{S:related}. The paper
is concluded in Sec.~\ref{S:concl}.


\section{Concerns in datatype transformation}
\label{S:concerns}

\vspace{-22\in}

\noindent
The central contribution of the present paper is a simple,
well-defined, and `editing-complete' suite of operators for datatype
transformations. Before we embark on this suite, we identify the
concerns addressed by our approach:
\begin{itemize}
\item Datatype transformations via scripting or interactive tool support.
\item Well-defined primitives for datatype transformations.
\item Generic meta-programming for conciseness of datatype transformations.
\item Flexible means of referring to fragments of interest in datatype transformations.
\end{itemize}
We will now discuss these concerns in some depth.


\mysubsection{Scripting vs.\ interactive tool support}

From the point of view of a programmer, datatype transformations
should be founded on intuitive scenarios for adaptation. To actually
perform (datatype) transformations, there are two modes of
operation. The first mode is \emph{scripting}: the programmer encodes
the desired transformation as an expression over basic or higher-level
operators. The second mode is \emph{interactive} transformation based
on a corresponding GUI. The benefits of an interactive tool are rather
obvious. Such a tool is useful to issue a transformation on the basis
of an operator-specific dialogue, and to provide a tailored list of
options for transformations that make sense in a given context. A
crucial benefit of interactive transformation is that the GUI can be
used to provide feedback to the programmer: Which locations were
changed? Where is the programmer's attention needed to complete the
issued transformation scenario? The apparent benefits of scripting
such as the opportunities to revise transformations and to replay them
can be also integrated into an interactive setting.

\smallskip

\begin{figure}[t]
\begin{center}
\resizebox{.7\textwidth}{!}{\includegraphics{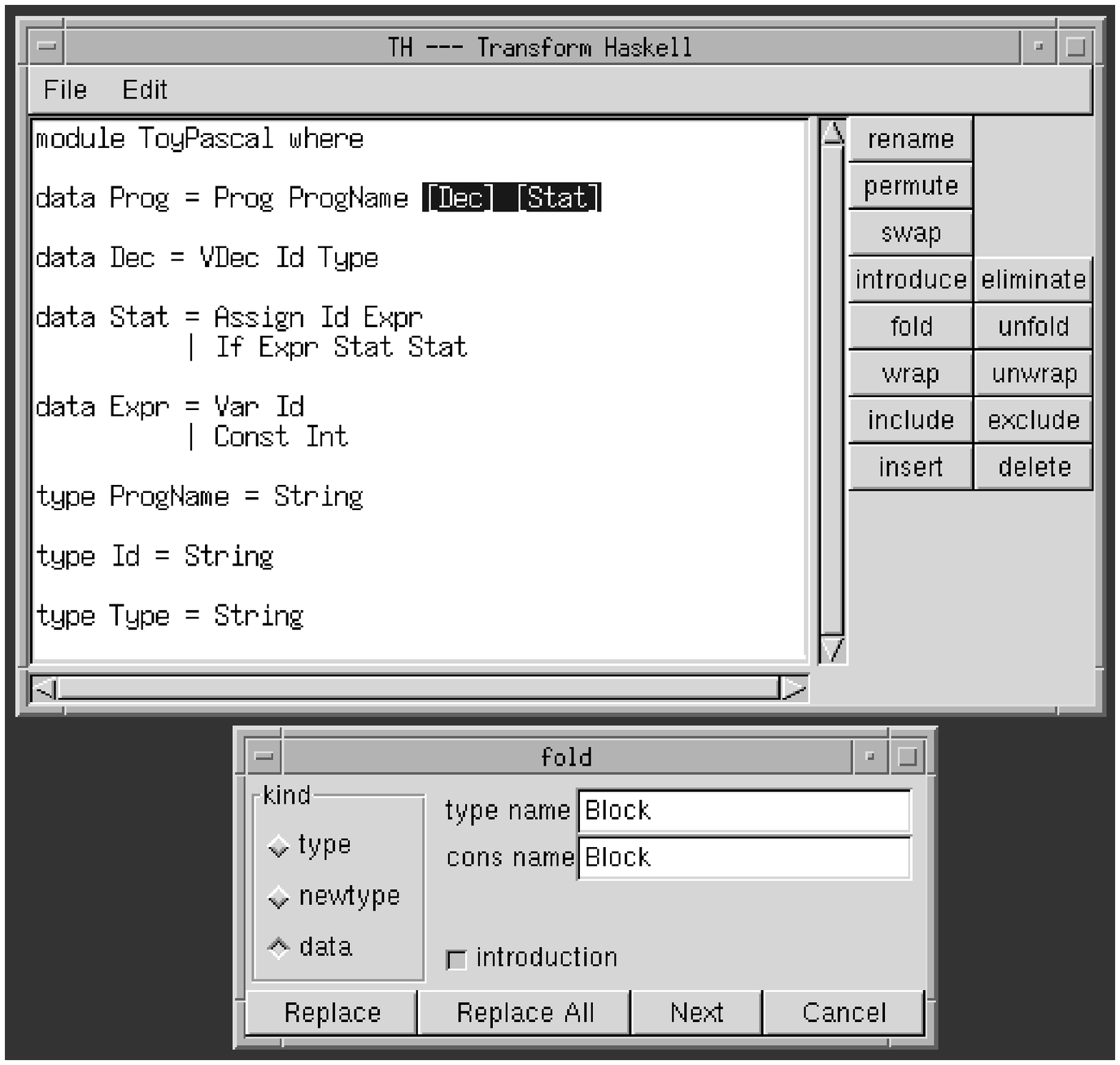}}
\end{center}
\vspace{-42\in}
\caption{A snapshot related to the interactive treatment of the introductory example}
\label{F:fdt}
\vspace{20\in}
\end{figure}

\noindent
In Fig.~\ref{F:fdt}, we illustrate the interactive treatment of the
introductory example using our prototypical tool
\emph{TH}~---~\emph{T}ransform \emph{H}askell. As the snapshot
indicates, we use a designated \emph{fold} dialogue to perform the
extraction of the piece of syntax. (Folding is the basic
transformation underlying extraction.) This dialogue combines several
transformation steps and side conditions in a convenient way. The
figure shows the following situation. The user has selected two
consecutive types ``\texttt{[Dec]} \texttt{[Stat]}'' and initiated the
\emph{fold} dialogue. The user has also typed in ``Block'' in the
``type name'' field.  The introduction check-box is marked
automatically since the given type name does not yet exist. The user
has also selected the ``kind'' radio-button to be ``data'' and filled
in ``Block'' in the ``cons name'' field. After this, the user would
press ``Replace'' to make the change.  If there had been more than one
occurrence, the user could replace them all with ``Replace All'', or
step through all occurrences with ``Next'', and replace only specific
ones with ``Replace'' as with ordinary \emph{find and replace} in text
editors.

\smallskip

\newpage

\noindent
Here is an open-ended list of further common transformation scenarios:
\begin{itemize}
\item Renaming type and constructor names.
\item Permuting type arguments and constructor components.
\item The dual of extracting datatypes, i.e., inlining datatypes.
\item Including a constructor declaration together with associated functionality.
\item Excluding a constructor declaration together with associated
functionality.
\item Inserting a constructor component together with associated functionality.
\item Deleting a constructor component together with associated functionality. 
\end{itemize}


\mysubsection{Well-defined transformation primitives}

The core asset of our framework is a suite of \emph{basic} operators,
which can be either used as is, or they can be completed into more
complex, \emph{compound} transformations. In the design of this suite,
we reuse design experience from a related effort on grammar
adaptation~\cite{Laemmel01-FME}. Indeed, there is an obvious affinity
of grammar transformations and datatype transformations.  A
challenging problem that we did not need to address in this previous
work, is the completion of datatype transformations to apply to entire
(functional) programs in which evolving datatypes reside.

\smallskip

\noindent
We list the required properties of our basic transformation operators:
\begin{description}
\item[Correctness] Mostly, we insist on `structure preservation', that
is, the resulting datatype is of the same shape as the original
datatype. This is enforced by the pre- and postconditions of the
operators.
\item[Completeness] The operators are `editing-complete', that is,
they capture all scenarios of datatype evolution that are otherwise
performed by plain text editors. Semantics-preserving adaptations are
defined in terms of disciplined primitives.
\item[Orthogonality] The operators inhabit well-defined,
non-overlapping roles. Higher-level scenarios for interactive
transformation are derivable. Operators for datatype transformations
are complementary to expression-level transformations.
\item[Locality] The basic operators operate on small code locations as
opposed to `global' or `exhaustive' operators, which iterate over the
entire program. Note that some operators are necessarily exhaustive,
e.g., an operator to rename a type name.
\item[Implementability] The operators are implemented as syntactical
transformations that are constrained by simple analyses to check for
pre- and postconditions, but which otherwise do not necessitate any
offline reasoning.
\item[Universality] While the present paper focuses on datatype
transformations, the principles that are embodied by our operators are
universal in the sense that they also apply to other abstractions than
datatypes, e.g., functions or modules.
\end{description}

\noindent
We do not list these properties to announce a formal treatment. This
would be very challenging as we opt for the complex language setup of
Haskell. The above properties provide merely a design rationale. A
formal approach is an important subject for future work, but it does
not contribute anything to the narrow goal of the present paper: to
compile an inventory of the basic roles in datatype transformation.


\mysubsection{Generic meta-programming}

We implement transformation operators and compound meta-programs in
Haskell. We reuse a publicly available abstract syntax for
Haskell.\footnote{The used abstract syntax is part of the Haskell Core
Libraries~---~in the haskell-src package.} We rely on generic
programming techniques to perform meta-programming on the non-trivial
Haskell syntax in Haskell. We use the
Strafunski-style\footnote{\url{http://www.cs.vu.nl/Strafunski/}} of
generic programming that allows us to complete functions on specific
syntactical sorts into generic traversals that process subterms of the
specific sorts accordingly. This style of meta-programming is known to
be very concise because one only provides functionality for the types
and constructors that are immediately relevant for the given problem.

\smallskip

\noindent
All our datatype transformations are of type \emph{Trafo} which is
defined as follows:

{\small
\smallskip\noindent{\footnotesize\parbox{.984\textwidth}{\input{snip/Trafo.math}\hfill\vspace{-82\in}}}\smallskip
}

\smallskip

\noindent
That is, a datatype transformation is a partial function on
\emph{HsModule}~---~the abstract syntactical domain for Haskell
modules. Partiality is expressed by means of the $\mathit{Maybe}$ type
constructor that wraps the result type. Partially is needed to model
side conditions.

\smallskip

\begin{figure}[t]\framebox{\parbox{.96\textwidth}{{\scriptsize\input{%
snip/renameTypeId.math}}\hfill\vspace{-75\in}}}%
\caption{%
Specification of the replacement operation underlying renaming of type names}
\label{F:%
renameTypeId}
\end{figure}

\noindent
In Fig.~\ref{F:renameTypeId}, we illustrate generic meta-programming
by giving the definition of a simple operator for replacing type
names. The specification formalises the fact that type names can occur
in two kinds of locations: either on a declaration site, when we
declare the type, or on a using site, when we refer to the type in a
type expression. So we need to synthesise a transformation which pays
special attention to the syntactical domains for declaring and using
sites. Indeed, in the figure, there are \emph{two} type-specific
`ad-hoc' cases which customise the identity function \emph{idTP}. In
the given context, we choose the traversal scheme \emph{full\_tdTP}
for `\emph{full} \emph{t}op-\emph{d}own traversal in
\emph{T}ype-\emph{P}reserving manner'. This way, we will reach each
node in the input tree to transform type names on declaring and using
sites. The operator \emph{replaceTypeId}, by itself, is a total
function.  (So the \emph{Maybe} in its type is not really needed
here.)  Partiality would be an issue if we derived an operator for
renaming type names. This necessitates adding a side condition to
insist on a fresh new name.

\mysubsection{Means of referring to fragments of interest}

Both the basic operators for datatype transformation but also actual
transformation scenarios in scripts or in interactive sessions need to
refer to program fragments of interest. Recall our introductory
example.  Extracting a type necessitates referring to the constructor
components that are meant to constitute the new type. In our
framework, we use three ways to refer to fragments of interest:
\begin{description}
\item[Focus markers on subterms] This approach is particularly suited
for interactive transformations. Here, relevant fragments can be
directly marked. In Fig.~\ref{F:Focus}, we extend Haskell's abstract
syntax to include term constructors for focusing on relevant fragments
in datatype transformations.  That is, we are prepared to focus on
names of types, on type expressions, and on lists of constructor
components.
\item[Selectors of subterms] This approach is particularly suited for
scripting transformations. Selectors for Haskell's type expressions
are defined in Fig.~\ref{F:TypeSel}. The three forms of \emph{TypeSel}
represent the three kinds of declarations that involve types. The
helper $\mathit{TypeSel}'$ allows to select any part of a given type
expression.
\newpage
\item[Predicates on subterms] Such predicates typically constrain the
type of a term or the top-level pattern. This approach is particularly
suited for the repeated application of a transformation to different
focuses that match a given predicate.
\end{description}

\smallskip

\begin{figure}[t]\framebox{\parbox{.96\textwidth}{{\scriptsize\input{%
snip/Focus.math}}\hfill\vspace{-75\in}}}%
\caption{%
Kinds of focus for datatype transformation}
\label{F:%
Focus}
\end{figure}

\begin{figure}[t]\framebox{\parbox{.96\textwidth}{{\scriptsize\input{%
snip/TypeSel.math}}\hfill\vspace{-75\in}}}%
\caption{%
Selectors that refer to type expressions, and others}
\label{F:%
TypeSel}
\end{figure}

\noindent
There are ways to mediate between these different ways of referring to
subterms. For example. given a term with a focus marker on a type
expression, one can compute the selector that refers to the focused
subterm.  Given a predicate on type expressions, one can compute the
list of all selectors so that an operator that is defined on selectors
can be used with predicates as well. Finally, given a selector, one
can also add the corresponding focus marker in the input at hand.


\vspace{-32\in}

\section{Basic operators for datatype transformation}
\label{S:suite}

\vspace{-22\in}

\noindent
We will now describe the themes that constitute our operator suite:
\begin{itemize}
\item Renaming type and constructor names.
\item Permutation of type parameters and constructor components.
\item Swapping types on use sites.
\item Introduction vs.\ elimination of type declarations.
\item Folding vs.\ unfolding of type declarations.
\item Wrapping vs.\ unwrapping of constructor components.
\item Inclusion vs.\ exclusion of entire constructor declarations.
\item Insertion vs.\ deletion of constructor components.
\end{itemize}
As this list makes clear, we group an operator with its inverse such
as in ``folding vs.\ unfolding'', unless the operator can be used to
inverse itself. This is the case for renaming, permutation, and
swapping. The operators from the first six groups are (almost)
\emph{structure-preserving}. The last two groups deal with
\emph{structure-extending} and \emph{-reducing transformations}.  We
will now explain the operators in detail including illustrative
examples. We will only explain the effect of the operators on datatype
declarations while we postpone lifting the operators to the level of
complete programs until Sec.~\ref{S:meeting}.

\mysubsection{Renaming and permutation}

Let us start with the simplest datatype refactorings one can think of.
These are transformations to consistently rename type or constructor
names, and to permute parameters of type and constructor
declarations. In Fig.~\ref{F:rename}, a simple example is illustrated.
We rename the type name $\mathit{ConsList}$, the constructor names
\emph{Nil} and \emph{Cons}, and we permute the two parameter positions
of \emph{Cons}. The resulting datatype specifies a $\mathit{SnocList}$ as
opposed to the $\mathit{ConsList}$ before.

\smallskip

\begin{figure}[t!]\framebox{\parbox{.96\textwidth}{{\scriptsize\input{%
snip/RenamePermuteSample.math}}\hfill\vspace{-75\in}}}%
\caption{%
Illustration of renaming and permutation}
\label{F:%
rename}
\end{figure}

\begin{figure}[t]\framebox{\parbox{.96\textwidth}{{\scriptsize\input{%
snip/RenamePermute.math}}\hfill\vspace{-75\in}}}%
\caption{%
Operators for renaming and parameter permutation}
\label{F:%
RenamePermute}
\end{figure}

\begin{figure}[t]\framebox{\parbox{.96\textwidth}{{\scriptsize\input{%
snip/RenamePermuteEncoding.math}}\hfill\vspace{-75\in}}}%
\caption{%
Script for the scenario in Fig.~\ref{F:rename}}
\label{F:%
rename-script}
\end{figure}

\noindent
In Fig.~\ref{F:RenamePermute}, we declare the operators for renaming
names and permuting parameter lists. In Fig.~\ref{F:rename-script}, we
include the script that encodes the
$\mathit{ConsList}$-to-$\mathit{SnocList}$ sample as a sequence of
basic renaming and permuting transformations. To this end, we assume a
sequential composition operator \emph{seqTrafo} for datatype
transformations. (In the script, \emph{seqTrafo} is used as an infix
operator `\emph{seqTrafo}`.)

\mysubsection{Introduction vs.\ elimination}

The next group of operators deals with the introduction and
elimination of type declarations (see Fig.~\ref{F:IntroElim}).
Introduction means that the supplied types are added while their names
must not be in use in the given program. Elimination means that the
referenced types are removed while their names must not be referred to
anymore in the resulting program. The two operators take \emph{lists}
of types as opposed to single ones because types can often only be
introduced and eliminated in groups, say mutually recursive systems of
datatypes. All kinds of type declarations make sense in this context:
aliases, newtypes, and proper datatypes. The operators for
introduction and elimination are often essential in compound
transformations. This will be illustrated below when we reconstruct
the introductory example in full detail (see
Sec.~\ref{S:reconstruct}).

\begin{figure}[t!]\framebox{\parbox{.96\textwidth}{{\scriptsize\input{%
snip/IntroElim.math}}\hfill\vspace{-75\in}}}%
\caption{%
Operators for introduction and elimination of datatypes}
\label{F:%
IntroElim}
\end{figure}

\mysubsection{Folding vs.\ unfolding}

Instantiating the folklore notions of unfolding and folding for
datatypes basically means to replace a type name by its definition and
vice versa. Extra provisions are needed for parameterised
datatypes. The prime usage scenarios for the two operators are the
following:
\begin{itemize}
\item \emph{extraction} = \emph{introduction} of a type followed by its \emph{folding}.
\item \emph{inlining} = \emph{unfolding} a type followed by its \emph{elimination}.
\end{itemize}
To give an example, the introductory example basically \emph{extracts}
the structure of imperative program blocks. To actually reconstruct
this example, we need a few more operators. So we postpone scripting
the example (see Sec.~\ref{S:reconstruct}).

\smallskip

\begin{figure}[t!]\framebox{\parbox{.96\textwidth}{{\scriptsize\input{%
snip/FoldUnfold.math}}\hfill\vspace{-75\in}}}%
\caption{%
Operators for folding and unfolding}
\label{F:%
FoldUnfold}
\end{figure}

\noindent
The operators for folding and unfolding are declared in
Fig.~\ref{F:FoldUnfold}, The operators make a strict assumption: the
type which is subject to folding or unfolding is necessarily a type
alias as opposed to a proper datatype.  This assumption simplifies the
treatment of the operators considerably since type aliases and their
definitions are equivalent by definition.  Extra operators for
so-called wrapping and unwrapping allow us to use proper datatypes
during folding and unfolding as well. This will be addressed below.
In the type of the \emph{foldAlias} operator, we do not just provide a
type name but also a list of type variables (cf.\ helper type
\emph{TypeHdr}). This is needed for parameterised datatypes, where we
want to specify how the free type variables in the selected type
expression map to the argument positions of the type alias.

\smallskip

\noindent
The preconditions for the operators are as follows. In the case of
\emph{foldAlias}, we need to check if the referenced type expression
and the right-hand side of the given alias declaration coincide. In
the case of unfolding, we need to check that the referenced type
expression corresponds to an application of a type alias.

\begin{figure}[t!]\framebox{\parbox{.96\textwidth}{{\scriptsize\input{%
snip/WrapUnwrap.math}}\hfill\vspace{-75\in}}}%
\caption{%
Operators for wrapping and unwrapping}
\label{F:%
WrapUnwrap}
\end{figure}

\begin{figure}[t]\framebox{\parbox{.96\textwidth}{{\scriptsize\input{%
snip/NormSample.math}}\hfill\vspace{-75\in}}}%
\caption{%
Illustration of wrapping, unwrapping, and extraction}
\label{F:%
NormSample}
\end{figure}

\mysubsection{Wrapping vs.\ unwrapping}
\label{S:reconstruct}

We will now consider operators that facilitate certain forms of
wrapping and unwrapping of datatype constructors (see
Fig.~\ref{F:WrapUnwrap}). There are operators for grouping and
ungrouping, that is, to turn consecutive constructor components into a
single component that is of a product type, and vice versa. There are
also operators to mediate between the different kinds of type
declarations, namely type aliases, newtypes and datatypes. This will
allow us to toggle the representation of datatypes in basic ways. As a
result, the normal forms assumed by other operators can be
established; recall, for example, the use of type aliases in folding
and unfolding. This separation of concerns serves orthogonality.

\smallskip

\begin{figure}[t]\framebox{\parbox{.96\textwidth}{{\scriptsize\input{%
snip/NormSampleEncoding.math}}\hfill\vspace{-75\in}}}%
\caption{%
Script for the scenario in Fig.~\ref{F:NormSample}}
\label{F:%
NormSampleEncoding}
\end{figure}

\noindent
In Fig.~\ref{F:NormSample}, we show the steps that implement the
introductory example. As one can see, we basically implement
extraction, but extra steps deal with grouping and ungrouping the two
components subject to extraction. Also, the extracted type should be a
proper datatype as opposed to a type alias (see transition from 3.\ to
4.). For completeness' sake, the transformation script is shown in
Fig.~\ref{F:NormSampleEncoding}. The script precisely captures the
steps that underly the interactive transformation in Fig.~\ref{F:fdt}.

\smallskip

\noindent
Some of the operators are not completely structure-preserving, that
is, strictly speaking, the structures of the datatypes before and
after transformation are not fully equivalent. For example, a newtype
and a datatype are semantically distinguished, even if the defining
constructor declaration is the very same. (This is because a
constructor of a datatype involves an extra lifting step in the
semantical domain, i.e., there is an extra `bottom' element.) The
operators for grouping and ungrouping also deviate from full structure
preservation.

\begin{figure}[t!]\framebox{\parbox{.96\textwidth}{{\scriptsize\input{%
snip/Match.snip}}\hfill\vspace{-75\in}}}%
\caption{%
Illustration of the generalisation of $\mathit{Maybe}$ to $\mathit{ConsList}$}
\label{F:%
Match}
\end{figure}

\mysubsection{Swapping types on use sites}

We will now deal with transformations that eliminate or establish type
distinctions by what we call \emph{swapping} types on use sites.  In
Fig.~\ref{F:Match}, we illustrate a typical application of
swapping. In the example, we want to generalise the standard datatype
$\mathit{Maybe}$ to allow for lists instead. In fact, we do not want
to change the general definition of the library datatype
$\mathit{Maybe}$, but we only want to change it on one use site (not
shown in the figure). This is where swapping helps: as an intermediate
step, we can replace $\mathit{Maybe}$ on the use site by a newly
introduced datatype $\mathit{Maybe}'$ with equivalent structure. The
figure illustrates how subsequent adaptations derive the
\emph{ConsList} datatype from the clone of the \emph{Maybe} datatype.
In particular, we add the boxed constructor component.

\begin{figure}[t!]\framebox{\parbox{.96\textwidth}{{\scriptsize\input{%
snip/Replace.math}}\hfill\vspace{-75\in}}}%
\caption{%
Operators for swapping types on use sites}
\label{F:%
Swap}
\end{figure}

\smallskip

\noindent
The swapping operators are declared in Fig.~\ref{F:Swap}. There is one
operator for type aliases and another for datatype declarations. In
the case of proper datatypes, one needs to match the constructors in
addition to just the names of the types. This is modelled by the
helper datatype \emph{DataUnifier}. The type of the operator
\emph{swapData} clarifies that we are prepared to process a list of
\emph{DataUnifier}s. This is necessary if we want to swap mutually
recursive systems of datatypes.

\begin{figure}[t!]\framebox{\parbox{.96\textwidth}{{\scriptsize\input{%
snip/IncludeExclude.math}}\hfill\vspace{-75\in}}}%
\caption{%
Operators for inclusion and exclusion of constructor declarations}
\label{F:%
IncludeExclude}
\end{figure}

\mysubsection{Inclusion vs.\ exclusion}

We now leave the ground of structure-preserving transformations. That
is, we will consider transformations where input and output datatypes
are not structurally equivalent. In fact, we consider certain ways to
extend or reduce the structure of the datatype. The first
couple of structure-extending and -reducing transformations is about
inclusion and exclusion of constructor declarations (see
Fig.~\ref{F:IncludeExclude}). These operators are only feasible for
proper datatypes and not for type aliases or newtypes. (This is because
a type alias involves no constructor at all, and a newtype is defined
in terms of precisely one constructor declaration.)

\smallskip

\begin{figure}[t!]\framebox{\parbox{.96\textwidth}{{\scriptsize\input{%
snip/IncludeSample.math}}\hfill\vspace{-75\in}}}%
\caption{%
Illustration of constructor inclusion}
\label{F:%
IncludeSample}
\end{figure}

\noindent
In Fig.~\ref{F:IncludeSample}, we show an example for constructor
inclusion. In fact, we just continue the introductory example to make
use of the extracted block structure in a language extension for
statement blocks. That is, we include a constructor application for
\emph{Stat} to capture \emph{Block} as another statement form. This
continuation of the introductory example amplifies the intended use of
our operator suite: for program evolution in the sense of 
datatype refactoring and adaptation.

\mysubsection{Insertion vs.\ deletion}
\label{S:insertDelete}
\label{S:maybe2list}

Inclusion and exclusion of constructor declarations is about the
\emph{branching} structure of datatypes. We will now discuss operators
that serve for the insertion or deletion of constructor
\emph{components} (see Fig.~\ref{F:InsertDelete}). Insertion of a
component $c$ into a constructor declaration $C\ c_1\ \cdots\ c_n$
proceeds as follows. Given the target position for the new component,
be it $i \leq n+1$, the new constructor declaration is simply of the
form $C\ c_1\ \cdots\ c_{i-1}\ c\ c_i\ \cdots\ c_n$. In general, $c$
might need to refer to type parameters of the affected
datatype. Deletion of a constructor declaration relies on the
identification of the obsolete component.

\smallskip

\begin{figure}[t!]\framebox{\parbox{.96\textwidth}{{\scriptsize\input{%
snip/InsertDelete.math}}\hfill\vspace{-75\in}}}%
\caption{%
Operators for insertion and deletion of constructor components}
\label{F:%
InsertDelete}
\end{figure}

\begin{figure}[t!]\framebox{\parbox{.96\textwidth}{{\scriptsize\input{%
snip/Maybe2List.math}}\hfill\vspace{-75\in}}}%
\caption{%
Illustration of component insertion and type swapping}
\label{F:%
Maybe2List}
\end{figure}

\noindent
In Fig.~\ref{F:Maybe2List}, we elaborate on the earlier example for
generalising `maybies' to lists (recall Fig.~\ref{F:Match}).  At the top
of Fig.~\ref{F:Maybe2List}, we see three datatypes
$\mathit{TransRel}$, $\mathit{Maybe}$, and $\mathit{ConsList}$. The
idea is indeed to replace $\mathit{Maybe}$ by $\mathit{ConsList}$ in
the using occurrence in $\mathit{TransRel}$. (That is, we want to
allow for a function from $a$ to a list of $a$s instead of a partial
function from $a$ to $a$.)  We call this adaptation a generalisation
because a list is more general than an optional. In the initial phase
of the generalisation of $\mathit{Maybe}$, we disconnect the relevant
occurrence of $\mathit{Maybe}$ in $\mathit{TransRel}$ from other
possible occurrences in the program. So we introduce a copy
$\mathit{Maybe}'$ of $\mathit{Maybe}$, and we perform type swapping so
that $\mathit{TransRel}$ refers to $\mathit{Maybe}'$ instead of the
`read-only' $\mathit{Maybe}$. Now we need to make $\mathit{Maybe}'$
structurally equivalent to $\mathit{ConsList}$. This amounts to adding
a recursive component to the second constructor
$\mathit{Just}'$. Then, we can again swap types to refer to
$\mathit{ConsList}$ in the co-domain of $\mathit{TransRel}$.


\section{Datatype transformation meets program transformation}
\label{S:meeting}

\vspace{-22\in}

\noindent
We will now re-iterate over the groups of operators to investigate
their impact on functional programs. It would be utterly complex to
formalise the link between datatype and program transformation. The
mere specification of the transformations is already intractable for a
publication because of its size, and the number of details. So we will
describe the implied program transformations informally while omitting
less interesting details.

\mysubsection{Renaming}

Type names only occur inside type declarations and type annotations.
So there is no need to adapt expressions or function declarations
except for their signatures, or the type annotations of
expressions. Constructor names can very well occur inside patterns and
expressions that contribute to function declarations. Renaming these
occurrences is completely straightforward.

\mysubsection{Permutation}

The permutation of type parameters does not necessitate any completion
at the level of function declarations. The permutation of constructor
components, however, needs to be realized in patterns and expressions
as well. This is particularly simple for pattern-match cases because
all components are matched by definition. Hence, we can directly
permute the sub-patterns in an affected constructor
pattern. Witnessing permutations of constructor components in
expression forms is slightly complicated by currying and higher-order
style. Instead of permuting components in possibly incomplete
constructor applications, we could first get access to all components
by `$\lambda$-pumping': given a constructor $C$ with say $n$ potential
components according to its declaration, we first replace $C$ by
$\lambda x_1\ \cdots\ x_n.\ C\ x_1\ \cdots\ x_n$ as justified by
$\eta$-conversion. Then, we witness the permutation by permuting the
arguments $x_1$, \ldots, $x_n$ in the pumped-up expression. In the
presence of a non-strict language with an evaluation order on
patterns, the permutation of constructor components might actually
change the behaviour of the program regarding termination. We neglect
this problem. We should also mention that it is debatable if the
described kind of $\eta$-conversion is really what the programmer
wants because it obscures the code.

\mysubsection{Introduction vs.\ elimination}

Introduction does not place any obligations on the functions defined
in the same program. In the case of elimination, we have to ensure
that the relevant types are not used by any function.  If we assume
that all function declarations are annotated by programmer-supplied or
inferred signatures, then the precondition for elimination can be
checked by looking at these signatures. There is an alternative
approach that does not rely on complete type annotations: we check
that no constructor of the relevant types is used.

\mysubsection{Folding vs.\ unfolding}

The restriction of folding and unfolding to type aliases guarantees
that these operators do not necessitate any adaptation of the function
declarations. This is simply because interchanging a type alias and
its definition is completely structure- and semantics-preserving, by
definition. This is extremely convenient: despite the crucial role of
the operators for folding and unfolding, they do not raise any issue
at the level of function declarations.

\mysubsection{Wrapping vs.\ unwrapping}

\emph{Grouping and ungrouping} These operators are handled using the
same overall approach as advocated for the permutation of constructor
components. That is, in patterns we witness grouping or ungrouping by
inserting or removing the enclosing ``(~\ldots~)''; in expressions, we
perform $\eta$-conversion to access the relevant components, and then
we group or ungroup them in the pumped-up constructor application.

\smallskip

\noindent
\emph{Mediation between newtypes and datatypes} These datatype
transformations do not imply any adaptations of the functions that
involve the datatype in question. (As we indicated earlier, the extra
bottom value of a datatype, when compared to a newtype, allows a
program to be `undefined' in one more way.)

\smallskip

\noindent
\emph{Newtype to alias migration} We simply remove all occurrences of
the associated constructor both in pattern and expression forms. We
require that the relevant newtype is not covered by any instance
declaration of some type class or constructor class. Otherwise, we had
to inline these members in a non-obvious way prior to the removal of
the constructor. If we neglected this issue, the resulting program
either becomes untypeable, or a different instance is applied
accidentally, which would be hazardous regarding semantics
preservation.

\smallskip

\noindent
\emph{Alias to newtype migration} This operator requires a non-trivial
treatment for function declarations. The crucial issue is how to know
the following:
\begin{itemize}
\item What expressions have to be wrapped with the newtype constructor?
\item In what patterns does the newtype constructor need to be stripped?
\end{itemize}
Our approach is as simple as possible. We observe that the new newtype
might be used in the declarations of other datatypes. The
corresponding patterns and expressions can be easily located and
adapted as in the case of permutation, grouping, and ungrouping
(recall $\eta$-conversion etc.). We also need to adapt function
declarations if their argument or result types are known to refer to
the relevant alias. This basically means that we need to access the
affected arguments and result expressions in all relevant equations to
unwrap the arguments and wrap the result expressions. These
adaptations are slightly complicated by the fact that the affected
type alias can occur in arbitrarily nested locations.

\medskip

\begin{figure}[t!]\framebox{\parbox{.96\textwidth}{{\scriptsize\input{%
snip/newtype.math}}\hfill\vspace{-75\in}}}%
\caption{%
Function adaptation triggered by alias-to-newtype migration}
\label{F:%
newtype}
\end{figure}

\noindent
In Fig.~\ref{F:newtype}, we illustrate the effect of the
\emph{alias2newtype} operator in the introductory example. We show the
top-level interpreter function that maps over the statements of the
program. (The program name and the declarations do not carry any
semantics here.) The type of the function \emph{run} exhibits that the
meaning of a program is a computation that involves a \emph{State} for
the program variables. The adapted version of \emph{run} refers to 
the extra constructor \emph{Block}, which resulted from extraction.

\begin{figure}[t!]\framebox{\parbox{.96\textwidth}{{\scriptsize\input{%
snip/ReplaceSample.math}}\hfill\vspace{-75\in}}}%
\caption{%
Function adaptation triggered by type swapping}
\label{F:%
SwapSample}
\end{figure}

\mysubsection{Swapping types on use sites}

This operator relies on the same techniques as
\emph{alias2newtype}. However, instead of wrapping and unwrapping a
constructor. We invoke conversion functions that mediate between the
two structurally equivalent types. These mediators merely map old to
new constructors and vice versa, and hence they are immediately
induced by the datatype transformation itself, namely by the
\emph{DataUnifier}s passed to the swap operator. This approach 
implies that we only perform very local changes. The program code
will still work on the old datatypes thanks to the mediators.

\smallskip

\noindent
The impact of swapping types at the function level is illustrated in
Fig.~\ref{F:SwapSample}. We deal with the initial steps of the
$\mathit{Maybe}$-to-$\mathit{ConsList}$ migration in
Fig.~\ref{F:Maybe2List}, where we replace the occurrence of
$\mathit{Maybe}$ within $\mathit{TransRel}$ by a structurally
equivalent $\mathit{Maybe}'$. We show an illustrative function
$\mathit{deadEnd}$ which performs a test if the given transition
relation allows for a transition in the presence of a given state
$a$. The adapted function $\mathit{deadEnd}$ refers to the conversion
function $\mathit{toMaybe}$ prior to performing pattern matching on
the obsolete $\mathit{Maybe}$ type.

\mysubsection{Inclusion vs.\ exclusion}

Intuitively, the inclusion of a constructor should be complemented by
the extension of all relevant case discriminations. This normally
means to add a pattern-match equation (or a case to a case expression)
for the new constructor. Dually, exclusion of a constructor should be
complemented by the removal of all pattern-match equations (or cases)
that refer to this constructor. In the case of added pattern-match
equations, we view the right-hand sides of these equations as a kind
of `hot spot' to be resolved by subsequent expression-level
transformations. To this end, we use ``undefined'', i.e., ``$\bot$'',
as a kind of to-do marker. Dually, in the case of removed
constructors, we also need to replace occurrences of the constructor
within expressions by ``$\bot$''. When using interactive tool support,
these to-do markers are useful to control further steps in a
transformation scenario.

\smallskip

\begin{figure}[t!]\framebox{\parbox{.96\textwidth}{{\scriptsize\input{%
snip/IncludeSample2.math}}\hfill\vspace{-75\in}}}%
\caption{%
Inclusion of a constructor declaration}
\label{F:%
IncludeSample2}
\end{figure}

\noindent
In Fig.~\ref{F:IncludeSample2}, we progress with our running example
of an interpreter for an imperative language. We illustrate the step
where blocks are turned into another form of statements. Hence, the
shown output program involves a new pattern-match equation that
interprets statement blocks. This added equation reflects that the
meaning of such blocks is as yet undefined, subject to subsequent
adaptations.

\mysubsection{Insertion vs.\ deletion}

Inserting a component into a declaration for a constructor $C$ means
that all patterns with $C$ as outermost constructor must be adapted to
neglect the added component, and all applications of $C$ must be
completed to include ``$\bot$'' for the added component. Dually,
deletion of a component from $C$ means that all applications of $C$
and all patterns with $C$ as outermost constructor need to be cleaned
up to project away the obsolete component. Any reference to a pattern
variable for the obsolete component is replaced by ``$\bot$''. As in
the case of permutation and others, $\eta$-conversion is needed to
actually get access to constructor components in expressions.

\smallskip

\begin{figure}[t]\framebox{\parbox{.96\textwidth}{{\scriptsize\input{%
snip/InsertSample.math}}\hfill\vspace{-75\in}}}%
\caption{%
Illustration of the insertion of a constructor component}
\label{F:%
InsertSample}
\end{figure}

\noindent
In Fig.~\ref{F:InsertSample}, the insertion of a constructor component
is illustrated by continuing the scenario from
Fig.~\ref{F:SwapSample}. The adapted equation of $\mathit{toMaybe}$
involves an extended pattern. As the don't care pattern ``\_''
indicates, the definition of \emph{toMaybe} does not make use of the
added component. In fact, the definition of the function
\emph{deadEnd} does not need to be adapted; it only tests for the
availability of a transition step. Normally, other functions will
start to rely on the richer pattern.


\section{Related work}
\label{S:related}

\vspace{-22\in}

\noindent
\emph{Transformational program development}
Formal program transformation~\cite{BD77} separates two concerns: the
development of an initial, maybe inefficient program the correctness
of which can easily be shown, and the stepwise derivation of a better
implementation in a semantics-preserving manner. Partsch's
textbook~\cite{Partsch90} describes the formal approach to this kind
of software development. Pettorossi and Proietti study typical
transformation rules (for functional and logic) programs
in~\cite{PP96}. Formal program transformation, in part, also addresses
datatype transformation~\cite{deRE98}, say data refinement. Here, one
gives different axiomatisations or implementations of an abstract
datatype which are then related by well-founded transformation steps.
This typically involves some amount of mathematical program
calculation. By contrast, we deliberately focus on the more
syntactical transformations that a programmer uses anyway to adapt
evolving programs.

\smallskip

\noindent
\emph{Database schema evolution}
There is a large body of research addressing the related problem of
database schema evolution~\cite{BKKK87} as relevant, for example, in
database re- and reverse engineering~\cite{HTJC93}. The schema
transformations themselves can be compared with our datatype
transformations only at a superficial level because of the different
formalisms involved. There exist formal frameworks for the definition
of schema transformations and various formalisms have been
investigated~\cite{McBP97}. An interesting aspect of database schema
evolution is that schema evolution necessitates a database instance
mapping~\cite{BCN92}. Compare this with the evolution of the datatypes
in a functional program. Here, the main concern is to update the
function declarations for compliance with the new datatypes.  It seems
that the instance mapping problem is a special case of the program
update problem.

\smallskip

\noindent
\emph{Refactoring}
The transformational approach to program evolution is nowadays called
refactoring~\cite{OpdykePhD,Fowler99}, but the idea is not
new~\cite{ABFP86,GN90}. Refactoring means to improve the structure of
code so that it becomes more comprehensible, maintainable, and
adaptable. Interactive refactoring tools are being studied and used
extensively in the object-oriented programming
context~\cite{Moore96,RBJ97}. Typical examples of \emph{functional}
program refactorings are described in~\cite{Laemmel00-SFP99}, e.g.,
the introduction of a monad in a non-monadic program. The precise
inhabitation of the refactoring notion for functional programming is
being addressed in a project at the University of Kent by Thompson and
Reinke; see~\cite{TR01}. There is also related work on type-safe
meta-programming in a functional context, e.g., by
Erwig~\cite{ER02}. Previous work did not specifically address datatype
transformations. The refactorings for object-oriented class structures
are not directly applicable because of the different structure and
semantics of classes vs.\ algebraic datatypes.

\smallskip

\noindent
\emph{Structure editing}
Support for interactive transformations can be seen as a
sophistication of structure editing~\cite{RT88,Koorn94,KS98}. This
link between transformation and editing is particularly appealing for
our ``syntactical'' transformations. Not surprisingly, concepts that
were developed for structure editing are related to our work. For
example, in~\cite{SM99}, primitives of structure editing are
identified based on the notion of focus to select subtrees, and on
navigation primitives left, right, up and down. Trees, subtrees and
paths are here defined as follows:\\
\smallskip\noindent{\footnotesize\parbox{.984\textwidth}{\input{snip/oegge.math}\hfill\vspace{-82\in}}}\smallskip

\noindent
The $t$ in a subtree $(p,t)$ is the currently selected tree and it is
between the left and right trees in the top layer (the head of the
$p$). This approach does not account for the heterogeneous character
of language syntaxes, but it shows that the fact if a focus resides
in a term can be encoded in types.


\section{Concluding remarks}
\label{S:concl}

\vspace{-22\in}

\noindent
\emph{Contribution} We identified the fundamental primitives for
datatype transformation. These operators are meant to support common
scenarios of program adaptation in functional programming, or other
settings where algebraic datatypes play a role. In fact, all the
identified operators are universal in the sense, that they are also
meaningful for other program abstractions than just datatypes, e.g.,
function declarations. We deliberately focused on adaptations of
datatypes because a vast body of previous work addressed fold/unfold
transformations for recursive functions.  Despite the focus on
datatype transformations, we had to consider program transformations
that are necessitated by the modification of datatypes. Regarding the
executable specification of the operator suite, we adhered to the
formula: meta-programs = object-programs = Haskell programs. We
employed generic functional programming in the interest of
conciseness. We also employed designated means of referring to
fragments of interest, e.g., a focus concept.

\smallskip

\noindent
\emph{Partial project failure}
We are confident that the identified operators are sufficient and
appropriate for actual datatype transformations. We have attempted to
complement this framework development by actual interactive tool
support. We initially thought that using Haskell for this interactive
tooling as well would be a good idea. Since the actual transformation
operators are implemented in Haskell anyway, and the interactive
dialogues need to cooperate with the operator framework to perform
analyses, Haskell indeed seems to be the obvious choice. To make a
long story short, there are many GUI libraries for Haskell, but none
of them is suitable for developing a sophisticated GUI for interactive
program transformation at the moment. It seems that environments for
interactive language tools would provide a better starting point,
e.g., environments based on attribute grammars~\cite{RT88,KS98}.

\smallskip

\noindent
\emph{Perspective}
To cover full Haskell, a few further operators would have to be added
to our suite, in particular, operators that support type and
constructor classes. We should also pay full attention to some
idiosyncrasies of Haskell; cf.\ refutable vs.\ irrefutable
patterns. Then, there are also transformation techniques that seem to
go beyond our notion of program evolution but it is interesting to
cover them anyway. We think of techniques like turning a system of
datatypes into functorial style, or threading a parameter through a
system of datatypes. The ultimate perspective for the presented work
is to integrate the datatype transformations into a complete,
well-founded, and user-friendly refactoring tool for functional
programming along the lines of Thompson's and Reinke's research
project~\cite{TR01}. Another perspective for our research is to
further pursue the intertwined character of datatype and program
transformations in the context of XML format and API evolution.


\vspace{-42\in}

\bibliographystyle{alpha}
\bibliography{paper}

\end{document}